Epitaxial Growth of VO$_2$ by Periodic Annealing

J. W. Tashman,[1] J. H. Lee,[1,2] H. Paik,[1] J. A. Moyer,[3,4] R. Misra,[5] J. A. Mundy,[6] T. Spila,[4] T. A. Merz,[1] J. Schubert,[7] D. A. Muller,[6,7] P. Schiffer,[3,4] and D. G. Schlom[1,8]

[1] Department of Materials Science and Engineering, Cornell University, Ithaca, New York 14853-1501, USA

[2] Neutron Science Division, Korea Atomic Energy Research Institute, Daejeon, Republic of Korea, 305-353

[3] Department of Physics, University of Illinois at Urbana-Champaign, Urbana, IL 61801, USA

[4] Frederick Seitz Materials Research Laboratory, University of Illinois at Urbana-Champaign, Urbana, IL 61801, USA

[5] Department of Physics and Materials Research Institute, Pennsylvania State University, University Park, Pennsylvania 16802, USA

[6] School of Applied and Engineering Physics, Cornell University, Ithaca, NY 14853, USA

[7] Peter Grünberg Institute, PGI 9-IT, JARA-FIT, Research Centre Jülich, D-52425 Jülich, Germany

[8] Kavli Institute at Cornell for Nanoscale Science, Ithaca, New York 14853, USA

Email: schlom@cornell.edu

We report the growth of ultrathin $VO_2$ films on rutile $TiO_2$ (001) substrates via reactive molecular-beam epitaxy. The films were formed by the cyclical deposition of amorphous vanadium and its subsequent oxidation and transformation to $VO_2$ via solid-phase epitaxy. Significant metal-insulator transitions were observed in films as thin as 2.3 nm, where a resistance change $\Delta R/R$ of 25 was measured. Low angle annular dark field scanning transmission electron microscopy was used in conjunction with electron energy loss spectroscopy to study the film/substrate interface and revealed the vanadium to be tetravalent and the titanium interdiffusion to be limited to 1.6 nm.

The huge metal-insulator transition (MIT) exhibited by $VO_2$ in the vicinity of room-temperature has made it a material of interest for uncooled microbolometer arrays,[1] gas sensing,[2] optical limiting,[3] and most recently MIT transistors.[4,5] In bulk single crystals this MIT occurs at a transition temperature ($T_c$) of 340 K and is accompanied by a change in structure from a high-temperature tetragonal form, to a low-temperature monoclinic form.[6] The change in resistivity through this transition in bulk $VO_2$ single crystals has been measured to be five orders of magnitude with a temperature hysteresis 0.5-1 K.[7] The change in resistivity in thick films (>100 nm) can be as high as four orders of magnitude,[8-10] but in thin films (<10 nm) is less than three orders of magnitude in all reports to date.[11-13]

While $VO_2$ presents an opportunity for emergent switching devices and sensors, its large carrier concentration (~$10^{22}$ cm$^{-3}$)[4,5] poses a serious challenge to using an applied electric field to traverse the MIT of a $VO_2$ channel in a field-effect transistor utilizing conventional solid-state dielectrics.[4] One approach to this challenge is to utilize organic ionic liquids as the gate dielectric. Ionic liquids can produce extremely high surface charge densities (~$10^{15}$ cm$^{-2}$), but the slow time response of their space-charge polarization mechanism makes them unlikely to be used in commercial electronics, and some studies question their efficacy as a means for electric-field control of MIT transistors.[4,14] Although one group reports the use of ionic liquids for gating $VO_2$,[5] other reports indicate that the changes in $VO_2$ conductivity arise from chemical effects (e.g., oxygen vacancies[4] or hydrogen doping[14]) rather than electric-field effects. An alternative approach to electric-field control of the MIT of $VO_2$ is to make the $VO_2$ channel layer of the MIT transistor have a thickness comparable to the Thomas-Fermi screening length of a conventional solid-state gate dielectric. To accomplish this goal, ultrathin, high quality films of $VO_2$ must be developed to decrease the total number of carriers in the $VO_2$ channel.[15]

In this paper we describe a process for the growth of ultrathin $VO_2$ films by reactive molecular-beam epitaxy (MBE) and show that they exhibit clear MITs in films as thin as 2.3 nm. We investigate the properties of these films with four-circle x-ray diffraction (XRD), low-angle annular dark field (LAADF) scanning transmission electron microscopy (STEM), electron energy loss spectroscopy (EELS), and electronic transport measurements. We also describe a standardized method for calculating the magnitude, hysteresis, and $T_c$ of these transitions using Savitsky-Golay smoothing derivatives.[16]

All films in this paper were grown by MBE in a Veeco Gen10. X-ray diffraction spectra were collected with a Rigaku Smartlab system utilizing Cu $K_{\alpha 1}$ radiation with a 220 Ge two-bounce incident-beam monochromator and a 220 Ge two-bounce diffraction side analyzer crystal. STEM images were taken with an FEI Tecnai $G^2$ F20. $VO_2$ film thicknesses were calculated with data from Rutherford backscattering spectrometry (RBS) assuming the calibration films had bulk $VO_2$ density. Electrical transport data was taken using the standard four-contact van der Pauw method in a Quantum Design Physics Property Measurement System (PPMS) with contacts made using gold wire and silver paint. All growth temperatures were measured using a thermocouple in the substrate cavity, but not in contact with the substrate. During growth the film was monitored using reflection high-energy electron diffraction (RHEED).

In this work we sought to produce high quality ultrathin $VO_2$ films by oxide MBE displaying MITs with $\Delta R/R$ values as large as possible. We began with the procedure described by Sambi *et al.* for producing epitaxial $VO_2$ thin films on $TiO_2$ (110).[17-21] The key aspect of this procedure was the deposition of 0.2-0.5 monolayers (ML) of amorphous vanadium metal at room temperature followed by a 2 minute anneal at 423 K in $7.5\times10^{-7}$-$1.5\times10^{-6}$ Torr of oxygen during which it transforms into an epitaxial VO$_2$ layer. One ML corresponds to 5.2×10$^{14}$ vanadium atoms / cm$^2$ for growth on TiO$_2$ (110).[21] Subsequent to the anneal in oxygen the sample was cooled to the original deposition temperature and the next 0.2-0.5 ML of amorphous vanadium metal was then deposited in vacuum. This cycle was repeated to build up epitaxial VO$_2$ films 3-5 ML thick.[17-21] This process is similar in nature to Al Cho's early GaAs films grown using "epitaxy by periodic annealing;"[22] epitaxy by periodic annealing has also been used to grow epitaxial SrTiO$_3$ on Si.[23,24] No electrical transport measurements on the VO$_2$ films made by this procedure were reported by Sambi *et al*.[17-21] Our films grown on TiO$_2$ (001) substrates by MBE following this procedure[17-21] were epitaxial, but did not exhibit an MIT. Only after utilizing ~ 80% pure distilled ozone[25] in place of molecular oxygen was an MIT observed. Post growth anneals in distilled ozone were also found to improve the transport properties of the epitaxial VO$_2$ films. Low hydrogen background pressure (less than 4×10$^{-9}$ Torr) was also important to producing films with good electrical transport properties. Altering the "epitaxy by periodic annealing" growth procedure of Sambi *et al.* to yield a high change in resistance ($\Delta R/R$) at the MIT led us to the following modified method.

Immediately prior to growth the TiO$_2$ (001) substrates were outgassed at 473 K in a background pressure of 1×10$^{-6}$ Torr of distilled ozone for 5 minutes *in situ*. Each substrate was subsequently cooled to 423 K before returning the ambient atmosphere to vacuum (8×10$^{-8}$ Torr). Upon reaching 395 K the growth procedure was initiated. Figure 1(a) illustrates the substrate temperature (the temperature plotted is the thermocouple temperature) during the cyclic growth procedure. The images shown correspond to the third cycle after initiation of growth on the bare TiO$_2$ (001) substrate. Figure 1(b) shows RHEED images along the [100] azimuth of TiO$_2$ and VO$_2$ taken at times corresponding to those labeled with arrows in Fig. 1(a). At point (1) in Fig. 1 the previous layer has recrystallized (the end of the second cycle), as seen in the corresponding RHEED image, and the sample is cool and ready for the deposition of the next layer. At point (2) vanadium was deposited at a flux of $4\times10^{13}$ atoms / (cm$^2$ · s) in vacuum for a time (12 seconds) corresponding to one formula unit of $VO_2$ for epitaxial growth of $VO_2$ (001) on $TiO_2$ (001). The corresponding vanadium dose is $4.82\times10^{14}$ atoms / cm$^2$. In the corresponding RHEED image the amorphous nature of the deposited vanadium layer can be seen. The weak diffraction visible in this image originates from the previously crystallized monolayer underlying the amorphous vanadium overlayer. Following the deposition of the amorphous vanadium formula unit, the substrate was heated to 475 K while the background distilled ozone pressure was simultaneously raised to $7\times10^{-7}$ Torr. It was important that the pressure reach its maximum value prior to the substrate temperature reaching 443 K. Subsequent to reaching 475 K, at point (3), the substrate was cooled in a background pressure of $7\times10^{-7}$ Torr of distilled ozone until reaching 405 K. Between points (2) and (4) the added monolayer of $VO_2$ film recrystallizes by solid-phase epitaxy. The ambient atmosphere was then returned to vacuum prior to repeating the process. At point (4), prior to the deposition of another monolayer, the crystal structure has recovered to that observed at point (1). This cyclic process was repeated four times to grow the initial four epitaxial monolayers of $VO_2$ on $TiO_2$ (001). In subsequent cycles the vanadium dose was changed to two formula units of $VO_2$ (a dose of $9.65\times10^{14}$ atoms / cm$^2$) to grow the remaining thickness of the epitaxial $VO_2$ (001) films.

Upon completion of the final growth cycle the substrate was allowed to cool as before, though the background distilled ozone pressure was instead increased to $1\times10^{-6}$ Torr prior to the substrate reaching 373 K. At that point the substrate was rapidly heated to 673 K at approximately 3 K / sec and then rapidly cooled, all in a background pressure of $1\times10^{-6}$ Torr of distilled ozone, until the substrate reached 405 K, at which point the ambient atmosphere was returned to vacuum. Figure 2 shows RHEED images along the [100] azimuth of $TiO_2$ and $VO_2$ taken before the anneal, Fig. 2(a), and after it, Fig. 2(b). After the anneal the intensity of the diffraction spots has increased relative to the background, indicating an improvement in the quality of the film. This final anneal was arrived upon by assessing the effect of different annealing temperatures and durations on the magnitude and sharpness of the MIT.

Figure 3 shows x-ray diffraction results for $VO_2$ films with thicknesses between 2.3 nm and 6.7 nm grown on $TiO_2$ (001). These three films were chosen for this study, from a number of samples with qualitatively similar properties, because they were grown consecutively using comparable growth conditions and because their thicknesses were corroborated by RBS measurements. In Fig. 3(a) two clear peaks at $2\Theta = 65.68°$ and $2\Theta = 62.72°$ are apparent in the scan of the thickest film (6.7 nm). The former is identified as 002 diffraction from $VO_2$ and the latter is indicative of the 002 diffraction from the tetragonal $TiO_2$ substrate. The lack of other peaks suggests that the film is *c*-axis oriented and phase pure. Figure 3(b) shows rocking curves for the thickest film in Fig. 3(a). A comparison of the full width at half maximum (FWHM) value of the 002 $VO_2$ film peak (0.0045°) with the 002 substrate peak (0.004°) suggests the film is of similar structural quality as the substrate. Figure 3(c) shows the same data in Fig. 3(b) on a linear scale to make it clear that the FWHM of the two curves are comparable. The arrows in Fig. 3(c) indicate the half-maximum of the XRD intensity.

STEM was used to interrogate the film microstructure and orientation relationship with the substrate. The close atomic numbers of vanadium and titanium prompted the use of low angle annular dark field STEM (LAADF-STEM) to obtain contrast between the film and substrate. Figure 4(a) shows the film/substrate interface of the 6.7 nm thick film shown in Fig. 3(a). STEM imaging revealed the films to be continuous, relatively smooth, and epitaxial. The epitaxial orientation relationship between the film and the substrate was determined to be (001) $VO_2$ ∥ (001) $TiO_2$ and [100] $VO_2$ ∥ [100] $TiO_2$, consistent with prior studies.[4,11] No second phases or other orientations of $VO_2$ were observed.

Electron energy loss spectroscopy (EELS) was used to quantify the titanium interdiffusion between the $VO_2$ film and the $TiO_2$ substrate as well as the valence state of the vanadium atoms. The vanadium $L_{2,3}$ edge is consistent with tetravalent vanadium[26] and, as seen in Fig. 4(b), analysis of the titanium $L_{2,3}$ edge shows titanium interdiffusion to be limited to 1.6 nm at the interface. We define the interdiffusion distance as the distance between the film interface (where Ti and V concentrations cross in Fig. 4(b)) and the point at which the Ti concentration crosses 0 %. The exceptionally low growth temperature, much lower than the 558 K – 696 K typically used to grow epitaxial $VO_2$ films,[4,5,11-13] is responsible for the limited interdiffusion. The limited interdiffusion and consistent tetravalent oxidation state of vanadium throughout the film make it possible for substantially thinner $VO_2$ films to exhibit MITs.[4,27,28]

The growth process described has yielded the thinnest films yet to show an MIT and the only $VO_2$ thin films grown by MBE to show an MIT.[13,29] Figure 5 shows the raw, unsmoothed measurements of the temperature dependence of resistivity for epitaxial $VO_2$ films with thicknesses between 2.3 nm and 6.7 nm. For each of these films resistivity measurements were made as the film was warmed and again as it was cooled at a rate of about 2 K / min. In Table I the magnitude of the resistivity change, the temperature range over which the MIT occurs, the hysteresis, and the transition temperature $T_c$ are given for epitaxial $VO_2$ films with thicknesses between 2.3 nm and 6.7 nm. These values were calculated using Savitzky-Golay smoothing numerical derivatives using the procedure described below and as shown graphically in Fig. 6 for the specific case of a 3.3 nm thick epitaxial VO$_2$ film.[16] Figure 6(a) shows the raw, unsmoothed measurement of the temperature dependence of resistivity for that film.

The midpoint of the transition ($T_{mid}$) was determined using the first derivative of a 5-point moving smooth utilizing a quadratic least-squares best-fit function to the data. The middle of the transition was defined to occur where the value of $d(\log(\rho))/dT$ was at its minimum. Figure 6(b) shows the first derivative and, with a circle, the minimum of the derivative. The starting and ending temperatures of the transition ($T_{start}$ and $T_{end}$) were determined by taking the second derivative of a 5-point moving smooth utilizing a cubic least-squares best-fit function to the data. The criteria that $|d^2(\log(\rho))/dT^2| < \varepsilon$ was utilized, with $\varepsilon = 5 \times 10^{-4}$. Figure 6(c) shows the second derivative and, with a square and a triangle respectively, the start and end of the transition.

The transition width was defined as the average of $|T_{start} - T_{end}|$, calculated using data collected during warming and cooling. The MIT transition temperature, $T_c$, was defined as the average of the two $T_{mid}$ values calculated from the same data. The hysteresis was defined as the difference between the two $T_{mid}$ values. The transition width was found to decrease monotonically with film thickness, while the hysteresis increased monotonically with film thickness.

In summary, we have developed a method for the growth of ultrathin films of epitaxial VO$_2$. Crucial to the electrical transport properties of these films, and their exhibiting clear MITs at average thicknesses as thin as 2.3 nm, are the low level of titanium interdiffusion and the consistent tetravalent oxidation state of vanadium.[4,27,28] These attributes are characteristic of this growth method, which has produced the thinnest films yet to show sharp, large magnitude MITs.

With such ultrathin films the challenges associated with using conventional solid-state gate dielectrics in VO$_2$ MIT field-effect transistors can begin to be addressed.

J.W.T., J.H.L., H.P., and D.G.S. gratefully acknowledge the financial support of ONR through award N00014-11-1-0665.  T.A.M. and D.A.M. acknowledge the National Science Foundation through the MRSEC program (Cornell Center for Materials Research, DMR-1120296).  This work made use of the electron microscopy facility of the Cornell Center for Materials Research (CCMR) with support from the National Science Foundation Materials Research Science and Engineering Centers (MRSEC) program (DMR 1120296) and NSF IMR-0417392. Julia A. Mundy acknowledges financial support from the Army Research Office in the form of a National Defense Science & Engineering Graduate Fellowship and from the National Science Foundation in the form of a graduate research fellowship.  This work was performed in part at the Cornell NanoScale Facility, a member of the National Nanotechnology Infrastructure Network, which is supported by the National Science Foundation (Grant ECCS-0335765).

Table I. MIT characteristics of the same films studied in Fig. 3(a).

| Thickness (nm) | $T_c$ (K) | $\Delta R/R$ (orders of magnitude) | Transition Width (K) | Hysteresis (K) |
|---|---|---|---|---|
| 2.3 | 304 | 1.4 | 34.0 | 9 |
| 3.3 | 280.5 | 2.3 | 22.5 | 10 |
| 6.7 | 286.8 | 2.7 | 14.8 | 19.5 |

**Figure Captions**

**FIG. 1**. (a) Illustration of the substrate temperature and background pressure of distilled ozone ($P_{O_3}$) used during the VO$_2$ growth cycle. (b) RHEED images taken along the [100] azimuth of TiO$_2$ and VO$_2$ during the growth cycle of a 6.7 nm thick epitaxial VO$_2$ film. In both, the indicated times correspond to: (1) prior to vanadium deposition, (2) immediately following vanadium deposition, (3) post anneal, and (4) at the beginning of the next cycle.

**FIG. 2**. RHEED images along the [100] azimuth of a 6.7 nm thick epitaxial VO$_2$ film prior to the final anneal (a) and after the final anneal (b). These results are on the same film studied in Fig. 1.

**FIG. 3**. (a) $\Theta$-$2\Theta$ and (b) rocking curve x-ray diffraction scans of a 6.7 nm thick epitaxial VO$_2$ film grown on TiO$_2$ (001). These results are on the same film studied in Figs. 1 and 2. $\Theta$-$2\Theta$ XRD scans of 3.3 nm and 2.3 nm thick epitaxial VO$_2$ films are also shown in (a). (c) Shows the rocking curves of substrate and film overlayed on a linear axis so it can be seen that they have comparable FWHMs. The arrows indicate the half-maximum of the XRD intensity

**FIG. 4.** (a) LAADF-STEM image of the 6.7 nm thick epitaxial VO$_2$ film shown in Fig. 3(a). (b) Vanadium and titanium EELS $L$-edge signals showing the extent of titanium and vanadium interdiffusion across the VO$_2$ / TiO$_2$ interface.

**FIG. 5**. Raw, unsmoothed measurements of the temperature dependence of resistivity for epitaxial VO$_2$ films with thicknesses between 2.3 nm and 6.7 nm. The three films measured are the same ones studied in Fig. 3(a).

**FIG. 6**. (a) The start, middle, and end of the metal insulator transition of a 3.3 nm thick epitaxial $VO_2$ film calculated using Savitzky-Golay smoothing derivatives. The curve in (a) is the raw, unsmoothed measurement of the temperature dependence of resistivity for that film. The legend in (a) applies to (b) and (c) as well. (b) shows the first derivative and (c) shows the second derivative. These results are on the same film studied in Fig. 3(a).

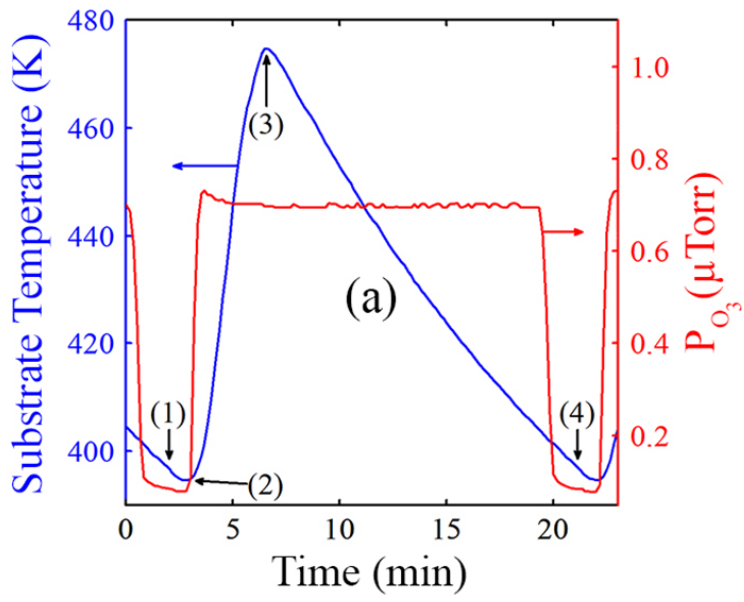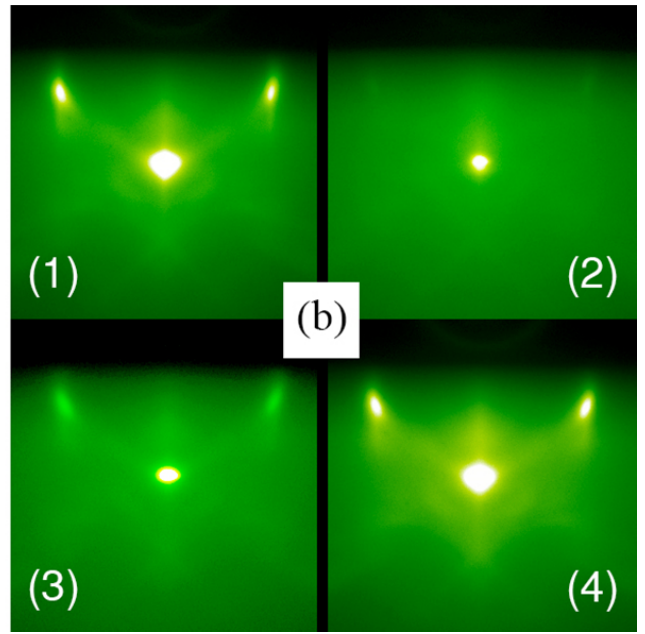

Figure 1

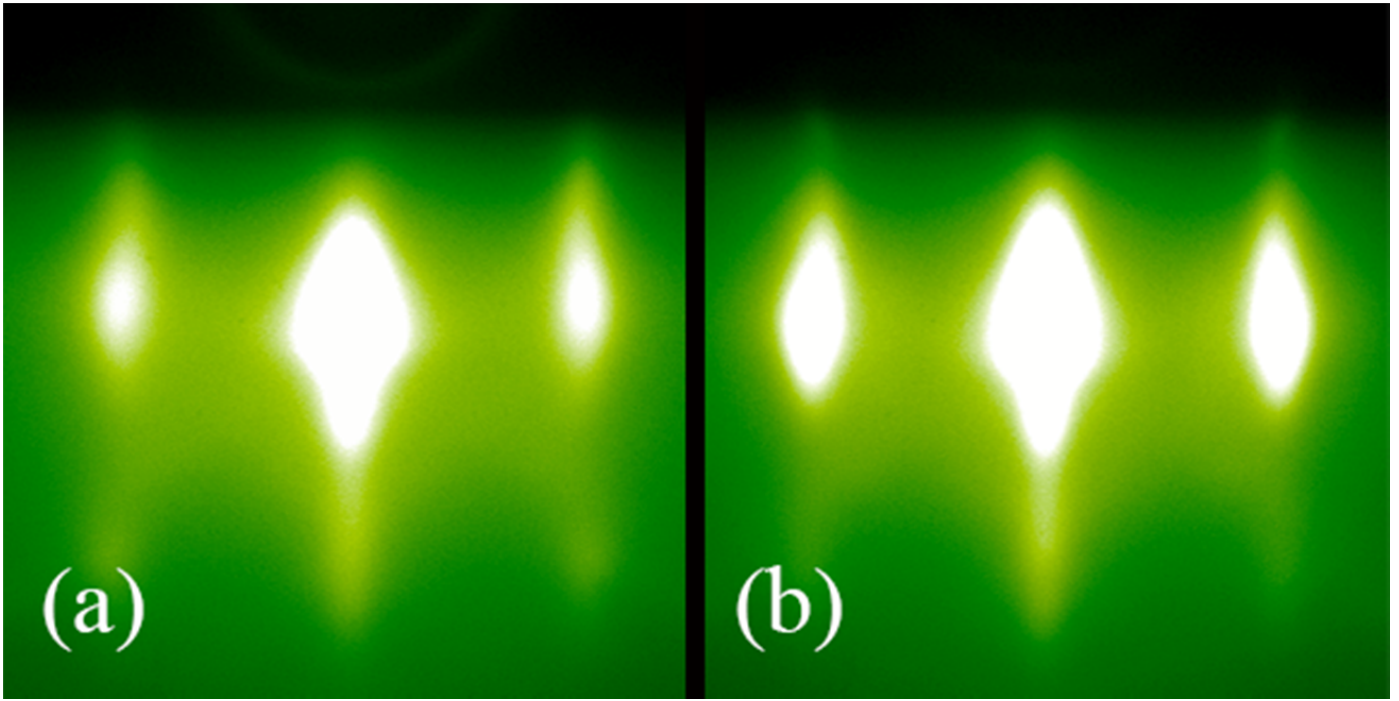

Figure 2

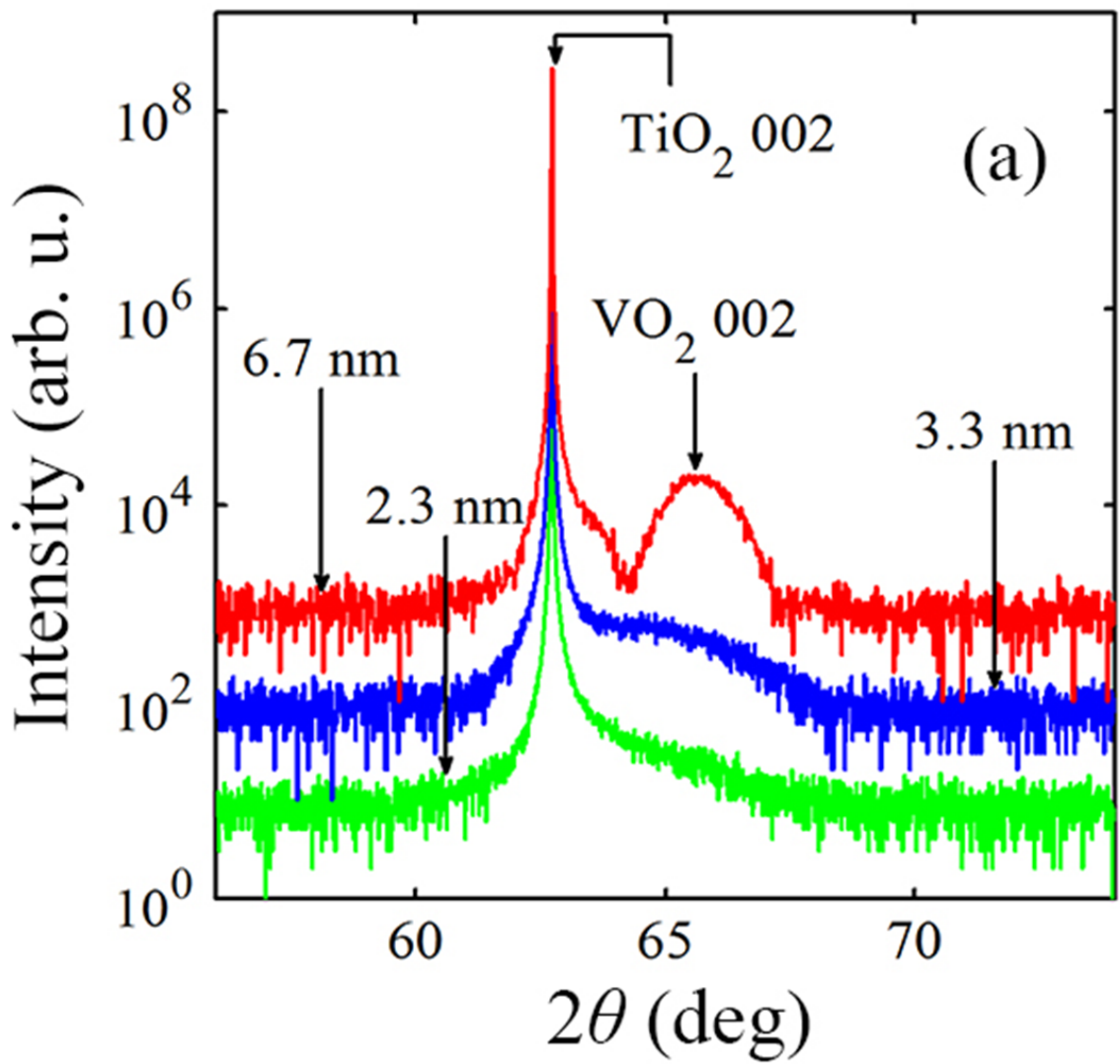

Figure 3a

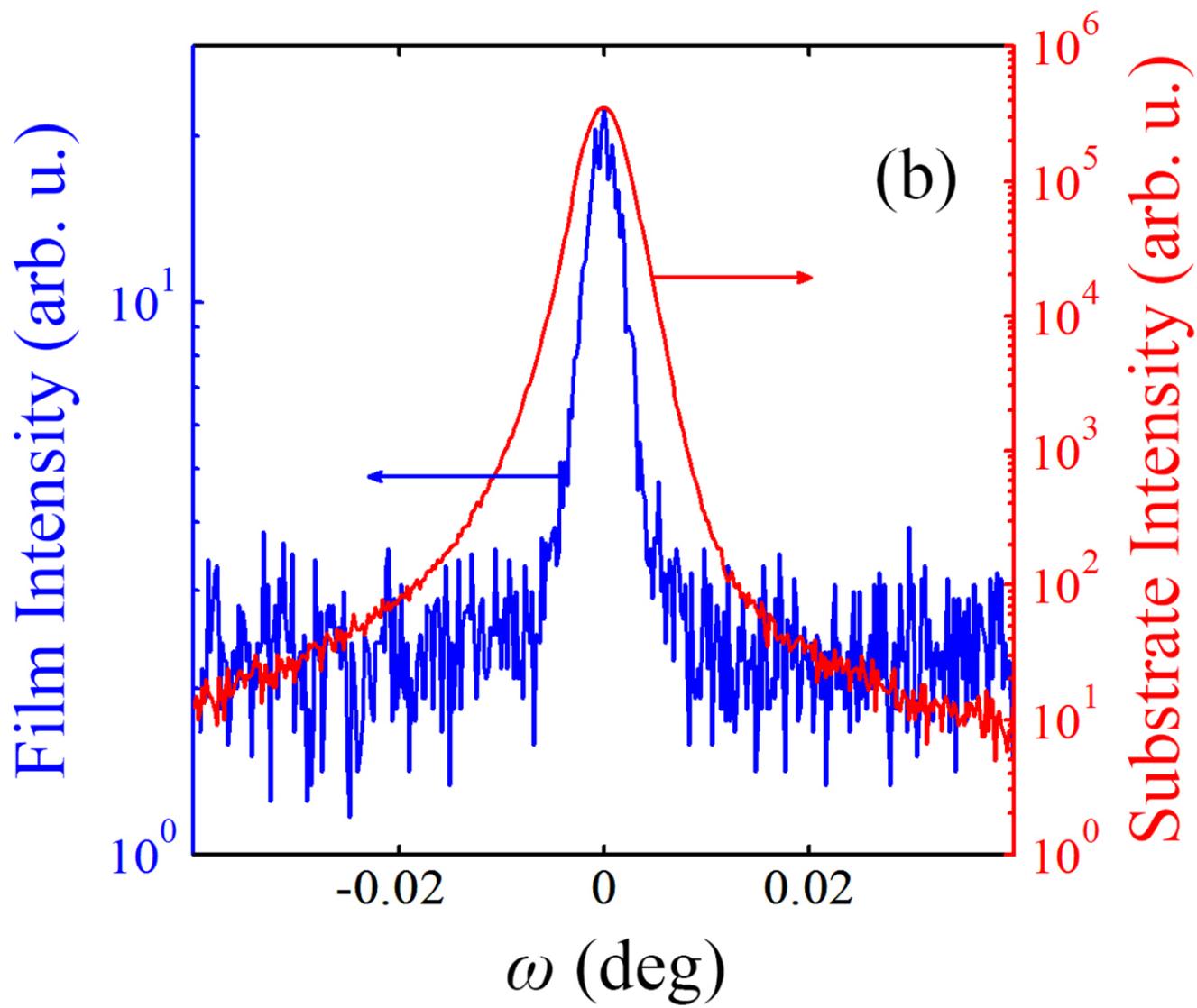

Figure 3b

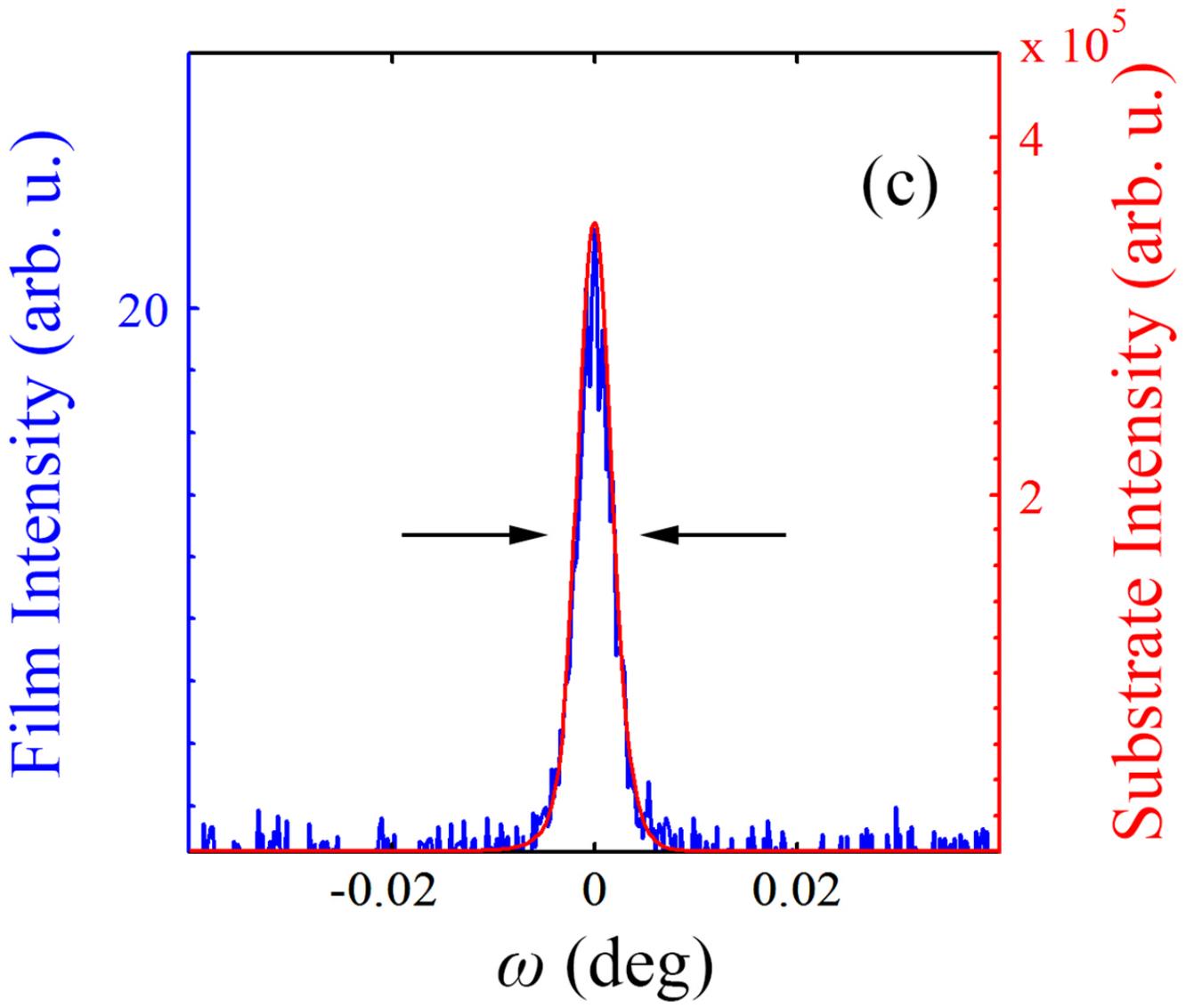

Figure 3c

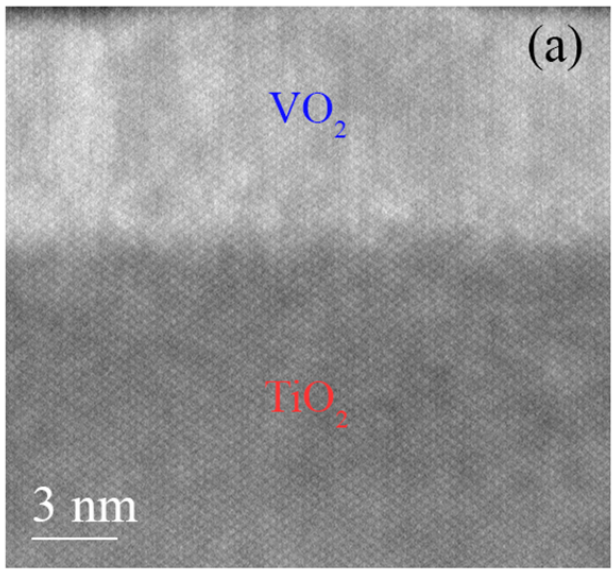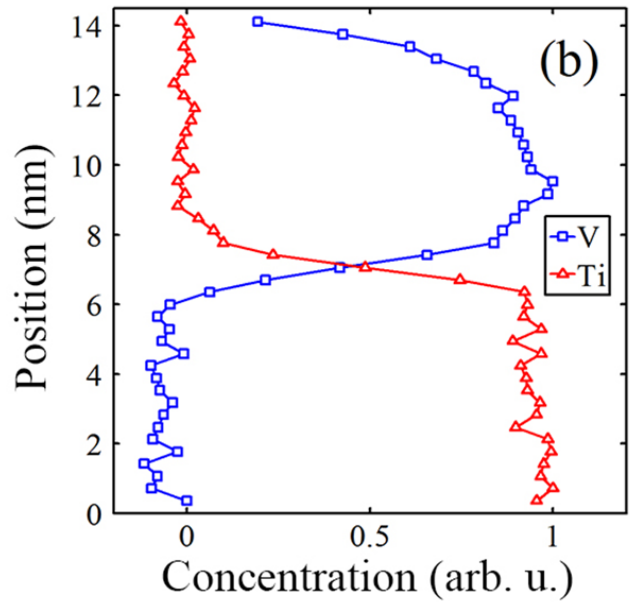

Figure 4

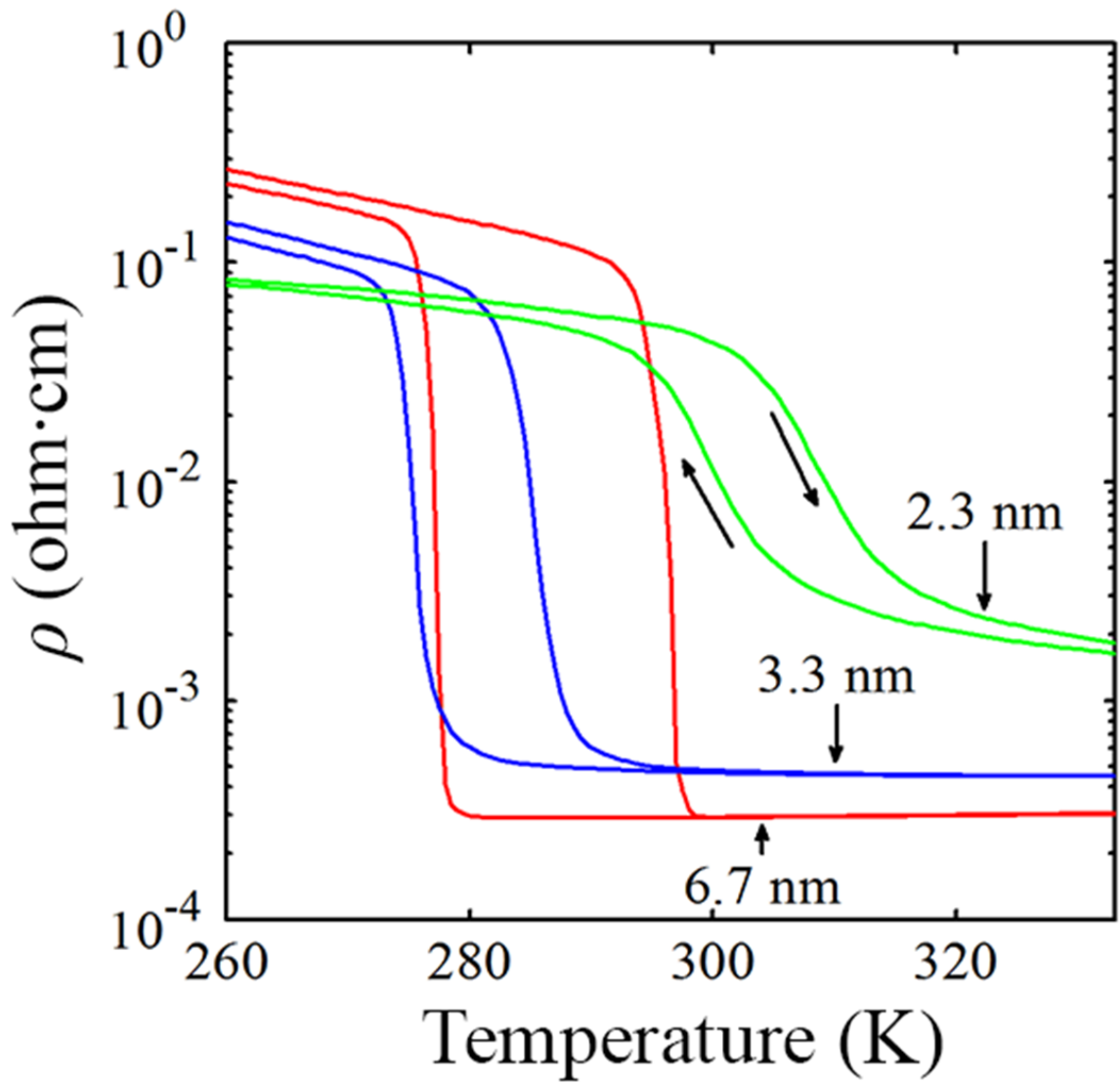

Figure 5

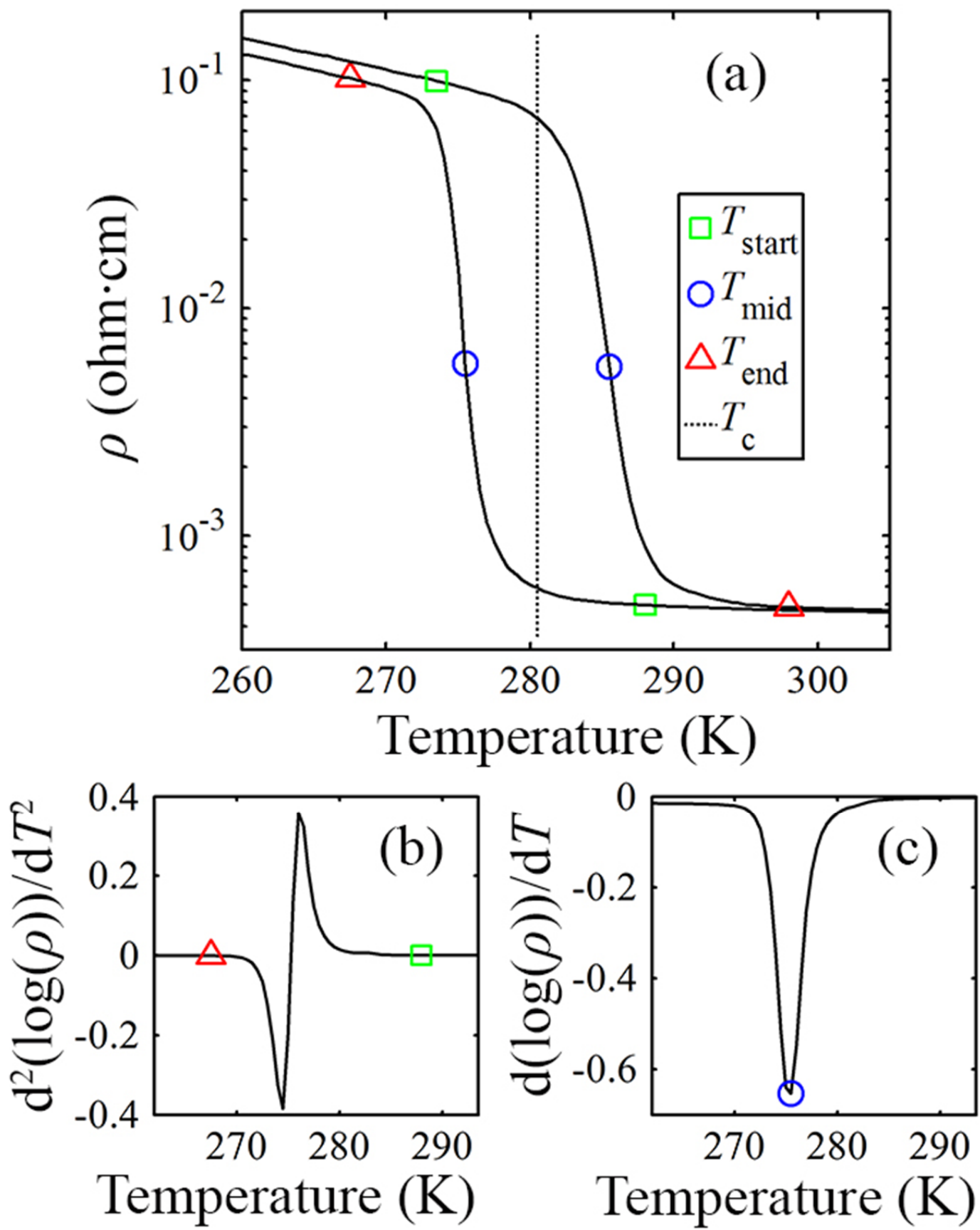

Figure 6